# HOW CELLS TIPTOE ON ADHESIVE SURFACES BEFORE STICKING

Anne Pierres[1,2,3,4], Anne-Marie Benoliel[1,2,3,4], Dominique Touchard[1,2,3,4], Pierre Bongrand[1,2,3,4]
[1]INSERM UMR600, [2]CNRS UMR6212, [3]Université de la Mediterranée, [4] APHM, Laboratoire "Adhésion et Inflammation", Campus de Luminy, Case 937, 13288 Marseille Cedex 09 FRANCE

*Condensed running title*: CELL MEMBRANE FITTING TO ADHESIVE SURFACES

*Keywords*: Cell adhesion, membrane undulations, interference reflection microscopy, monocyte, membrane fitting, membrane fluctuations

## ABSTRACT
Cell membranes are studded with protrusions that were thoroughly analyzed with electron microscopy. However, the nanometer-scale three-dimensional motions generated by cell membranes to fit the topography of foreign surfaces and initiate adhesion remain poorly understood. Here, we describe the dynamics of surface deformations displayed by monocytic cells bumping against fibronectin-coated surfaces. We observed membrane undulations with typically 5 nm amplitude and 5-10 second lifetime. Cell membranes behaved as independent units of micrometer size. Cells detected the presence of foreign surfaces at 50 nm separation, resulting in time-dependent amplification of membrane undulations. Molecular contact then ensued with apparent cell-membrane separation of 30-40 nm, and this distance steadily decreased during the following tens of seconds. Contact maturation was associated with in-plane egress of bulky molecules and robust membrane fluctuations. Thus, membrane undulations may be the major determinant of cell sensitivity to substrate topography, outcome of interaction and initial kinetics of contact extension

## INTRODUCTION

The contact between a cell and a foreign surface often triggers a variety of events such as cell attachment, spreading on the surface and alteration of behavioral patterns relative to survival, differentiation, migration or release of bioactive factors (1). During the last years, the biologically relevant phenomena occurring after the first minutes following cell-to-surface encounter were subjected to considerable scrutiny. Also, much effort was done to apply biophysical approaches to the analysis of the binding events occurring during the first seconds following contact (2-4). However, fewer studies were specifically focussed on the first minutes following cell-to-surface contact and preceding biologically relevant responses such as spreading, lamellipodium formation and migration, or activation of different functions. Interestingly, recent experiments performed with atomic force microscopy at the single cell level (5) suggest that cell adhesion to a surface may sequentially involve passive formation of a few receptor-ligand bonds followed a few minutes later by active cooperative events resulting in more than tenfold increase of adhesive strength. Thus, it is warranted to investigate the minute-scale period linking cell-substrate encounter to generation of an active response guided by specific recognition events involving cell membrane receptors and substrate molecular structures.

Numerous electron microscopical studies clearly demonstrated that cell membranes are studded with different membrane folds or protrusions such as microvilli, lamellipodia or filopodia (6-8). Cell adhesion is essentially mediated by membrane receptors of typically several tens of nanometer length (9), this is significantly shorter than the smallest protrusions covering the cell membrane, since microvilli are several tenths of micrometer long (8). Thus, formation of a substantial contact area on the order of several squared micrometers (10) requires nanometer-scale fitting of the cell topography to opposing surfaces. Indeed, while contact formation between a rough cell membrane and a smooth surface obviously requires a minimal smoothing of the cell surface, (11), intercellular adhesion also requires membrane adaptation (12).

While aforementioned requirements were fully illustrated with electron microscopy, little information is currently available on the dynamics of cell membrane at the nanometer scale. It has long been shown that the surface of erythrocytes (13) or deflated lipid vesicles display thermally driven oscillations, and much theoretical work was devoted to account for this phenomenon (14, 15). However, little information is available on the dynamics of nucleated cell membranes at the submicrometer level. Transverse oscillations of up to 20-30 nm amplitude were found on many nucleated cells such as fibroblasts, lymphocytes or monocytes (16-18). Other studies (8) revealed that leukocyte surface microvilli could withstand a pulling force of about 45 nN before forming a thin membrane tether. However, more work is needed to understand the relationship between membrane undulations and the early steps of contact formation and extension between a cell and an adhesive surface.

The purpose of the present work was to provide a real-time description of three-dimensional membrane deformations occurring during the first tens of seconds required to secure cell attachment to adhesive surfaces. We used interference reflection microscopy/reflection interference contrast microscopy (18-22) to image the initial interaction between human monocytic THP-1 cells and planar fibronectin-coated surfaces. Cell membranes displayed continuous transverse undulations of nanometer amplitude with a lateral correlation on the order of a micrometer. Cells seemed to detect the presence of the surface at about 50 nm separation as revealed by the behaviour of membrane fluctuations. Adhesion was not an immediate phenomenon, rather the cell-to-surface gap decreased by several tens of nanometers during the first few minutes following initial contact. This contact maturation occurred concomitantly with an egress of bulky components of the cell glycocalyx. Observation of cytoskelal elements suggested that actin acted at least partly as a stabilizing structure, and myosin was not detectable in contact areas.

## MATERIALS AND METHODS

**Cells and transfection** : We used the monocytic THP-1 line (23) maintained as previously described (24) in RPMI-1640 medium (Invitrogen) supplemented with 20 mM HEPES buffer, 10 % fetal calf serum (Invitrogen), 2mM L-glutamine, 50 U/ml penicillin and 50 µg/ml streptomycin. As previously checked with flow cytometry, these cells expressed typical mononuclear phagocyte markers including CD11b, CD18, CD29, CD32, CD43, CE45, CD64 and class I major histocompatibility complex molecules. Also they were shown to bind fibronectin-coated surfaces with CD49dCD29 (VLA-4) and CD49eCD29 (VLA-5) integrins (24).

THP-1 cells were transfected with pEGFP-actin (Clontech, BD Biosciences) using an AMAXA electroporator, AMAXA solution V and protocol V-01. EGFP-positive cells were selected with G418 selection and limiting dilution plating to isolate clones stably expressing EGFP-actin. For transfected cells 1mg/ml G-418 sulphate (GIBCO, Invitrogen Corporation) was added in culture medium.

**Fluorescent labelling and metabolic inhibitors :**
Cell overall shape was studied by monitoring cells suspended in culture medium containing 10 µg/ml fluorescein isothiocyanate with confocal microscopy.

In other experiments, cells were labelled with anti-CD43 monoclonal and processed for zenon labelling following the manufacturer (Molecular Probes) procedure. Mouse anti-CD29 (K20) and anti-CD45 were purchased from Beckman Coulter. Cytochalasin D (an inhibitor of microfilament assembly) and butanedione monoxime (a myosin inhibitor) were obtained from Sigma. Intracellular calcium visualization was achieved after labelling cells with Fluo-3-AM (Molecular Probes) as previously described (25)

**Chambers :** Custom-made flow chambers (Satim, Evenos, France) were Plexiglas blocks drilled with a rectangular cavity of $17 \times 6 \times 1$ mm$^3$ and cylindrical ducts for liquid entry and exit. The chamber floor was made of a glass coverslip maintained against the cavity with a steel plate.
Coverslips were cleaned by treatment with 98% sulphuric acid for 4 hours, washed overnight with tap running water, then distilled water and dried at 60° C for 30 min. They were then treated with 10 µg/ml fibronectin (Sigma) for 2 hours and rinsed in PBS.
Chambers were set on the stage of an Axiovert 135 inverted microscope (Zeiss) bearing a heating stage (TRZ 3700) set at 37°C. Interference reflection microscopy was performed with an antiflex objective (63 X magnification, 1.25 numerical aperture). Illumination was done with a 50 W mercury lamp using a H546 band-pass filter. Images were obtained with an Hamamatsu C4742-95-10 CCD camera yielding 10-bit accuracy. Pixel size was $125 \times 125$ nm$^2$.

**Spreading assays** were initiated by injecting 10µl of a concentrated cell suspension ($2 \times 10^6$ cells/ml) into the bottom of treated chambers containing normal culture medium. In a typical experiment, 300 sequential images were recorded with about 1 Hz frequency.

**Image processing** : Image processing was performed with a custom-made C++ Windows™ compatible software written in the laboratory. This allowed us to perform the following key steps as an improvement of a previously described and validated procedure (22)
- A first step consisted of performing a first order linear correction of brightness with respect to coordinates in order to minimize the illumination differences between the four corners of a $25 \times 25$ µm$^2$ area surrounding the cell. The same correction parameters were used throughout each series of 300 images, since they were essentially dependent on the cell position with respect to the microscope field.
- Secondly, the local intensity at each pixel was replaced with the average of intensities measured on the neighbouring 5×5 pixels. This was chosen to reduce the image noise sufficiently to decrease the height measurement error below about 1 nm, which was found necessary to process cell images. Note that the loss of lateral resolution did not result in dramatic loss of information as compared with the maximum resolution yielded with IRM/RICM.
- The last step consisted of deriving a relative membrane to substrate distance z from the brightness I with the formula :

$$z = (\lambda/4\pi) \text{ Arccosine } [(2I - I_m - I_M)/(I_m - I_M)] \qquad (1)$$

where $\lambda$ is the light wavelength, and $I_m$ and $I_M$ are respectively the minimum and maximum brightness corresponding to contact and ($\lambda/4$) separation. The interest of this formula is that estimates may be considered as fairly robust even if underlying assumptions are not fully warranted (22). Note also that the ambiguity linked to the arccosine transform is not a problem when we consider only a region close to the contact where the membrane-to-surface distance is expected to remain lower than about 100 nm. This assumption is supported by the absence of concentric fringes usually observed on the rim of molecular contact zones (21, 22, 26).

Parameters $I_m$ and $I_M$ were determined after examining a set of 300 images on the same microscope field. Corrected values were used for each image in order to account for the fluctuations of illumination intensity of a fixed empty area (the coefficient of variation was close to 1 %)

Note that this procedure was not expected to yield absolute distances. However, it must be recognized that there is a necessary arbitrariness in defining a nanometer-scale distance between surfaces coated with macromolecules of various shape.

The results presented in this paper were obtained after examining about 5,000 microscope fields of about $125 \times 125$ µm$^2$ each.

## RESULTS

**Cells adhere to adhesive surfaces and form growing contacts within less than one minute**.
Monocytic THP-1 cells were deposited on fibronectin-coated surfaces in a chamber set on a microscope stage. While these cells were fairly spherical in suspension, they displayed marked flattening at the micrometer scale within seconds following contact (Fig. 1A). Interference reflection microscopy/reflection interference contrast microscopy (IRM) was used to image the details of the cell surface in front of the substratum with nanometer vertical resolution. As shown on Fig. 1B-C, within a few tens of second following contact, a dark area appeared on the IRM image of the cell surface, revealing the formation of a contact zone with surface separation lower than several tens of nanometers. This was previously shown to reflect *bona fide* adhesion (22). Comparing the histograms of images obtained before and after adhesion (Fig. 1D) allowed us to define a threshold intensity (Arrow on Fig. 1D) to define contact zones and measure adhesive areas as was done in previous reports (10, 21, 27). The corresponding separation value obtained with Eq.1 was 34 nm. The kinetics of contact extension was then studied. As shown on Fig.1E, contact area first increased linearly with respect to time, and the growth rate was weakly dependent on surface concentration of adhesion molecules. This finding is consistent with the concept that contact extension should be essentially driven by cell movements, without any requirement for cooperative interactions between membrane receptors (5, 28).

As exemplified on Fig2A-I, growing contact zones exhibited two important features : i) the contact patches appeared fairly irregular and ii) some isolated patches sometimes appeared and disappeared, demonstrating the need to consider three-dimensional membrane topography in order to account for two-dimensional contact growth.

**Cell-substrate contact formation is associated with transverse membrane deformation and fluctuations of contact area superimposed on a continuous decrease of membrane to substrate distance.**
IRM images were processed to estimate cell-to-substrate distance within the 100 nm range. Images of cells falling on planar fibronectin-coated surfaces were recorded for periods of 5 minutes with 1Hz frequency. Three sequential color-coded maps of cell-surface distance are displayed on Fig.3 A-C, corresponding to the two-dimensional representations displayed on Fig. 2 A-C : clearly, contact formation was preceded with important cell surface deformations. These deformations were also displayed on Fig. 4 with a more intuitive 3D representation. Thus, contact initiation and extension were correlated with continuous membrane fluctuations. As shown on Fig. 5, these membrane undulations appeared concomitantly with two phenomena : i) contact area displayed marked fluctuations with about 0.1 Hz frequency, that were superimposed on average growth (Fig. 5A), and ii) during several tens of seconds following cell-to-surface contact, the relative molecular distance between interacting surfaces within previously defined contact zones steadily decreased by at least 30 nm (Fig. 5B). Thus membrane adhesion in a given point is not an all-or-none instantaneous process, but this process involves progressive narrowing of the cell-surface gap.

**Transverse membrane fluctuations may be described as a vertical motion of blocks of 0.6µm width and an effective diffusion coefficient of 1-3 $nm^2$/s.**

A conventional display of contact fluctuations is shown on Fig. 6A, emphasizing the concomitant expansion and retraction of contact in different zones. This two-dimensional view is a consequence of three-dimensional deformations of the cell membranes (Fig 6B) : different regions of the cell membrane facing the substratum displayed upward or downward displacements of 5-10 nm amplitude in different regions of the cell membrane during a 6-second period of time. These overall displacements were associated with membrane fluctuations that were revealed (Fig. 6C) by imaging the standard deviation of z coordinate on each pixel, during the same period of time and using higher image acquisition frequency. Then, to understand the rationale of cell shape control, the following analysis was performed

    i) we examined the randomness of membrane deformations by studying the time dependence of the squared displacements of individual surface pixels. Indeed, random walks result in a linear dependence of squared displacement on time, whereas a uniform motion would lead to a linear relationship between displacement and time. As exemplified on Fig. 7A, when the mean membrane approach to the surface was subtracted from invidual pixel displacement, the relationship between squared displacement and time was strikingly linear. This finding was obtained on nine separate cells and time intervals ranging between 6 and 80 seconds, immediately after contact, yielding an effective diffusion coefficient equal to half the slope of the curves (since the squared displacement during time t is expected to be equal to 2Dt). This ranged between about 1 and 3 $nm^2$ per second. Thus, membrane fluctuations separated by a 1-second period of time were not correlated.

    Further information was obtained by studying the temporal autocorrelation of membrane shape and fluctuation maps. As exemplified on Fig. 7B, the characteristic time for 50% decrease of correlation was about 5 seconds during contact initiation. However, the onset of adhesion resulted in progressive increase of correlation as a consequence of the formation of fixed adhesive zones (not shown). We also studied the time correlation of displacements (i.e. of the time derivative of membrane maps). No correlation could be found between sequential displacements (Fig. 7B). Together with Fig6A, this is consistent with the view that the characteristic duration of membrane fluctuations was significantly less that 1 second, but a longer correlation was generated by the formation of adhesive patches.

    ii) Then we examined the dependence of the amplitude of membrane fluctuations on z coordinate (i.e. distance to the adhesive substrate). Indeed, passive surface undulations of soft vesicles have long been studied (14, 15), and the squared amplitude of undulations should be decreased by a neighbouring surface in proportion to the squared distance (29). The mean correlation coefficient between distance to the fibronectin-coated surface and undulation amplitude calculated on a series of 16 different images was 0.101 (±0.358 SD). The significance of this unexpectedly low correlation was explained by plotting fluctuations (expressed as a standard deviation) versus distance. As exemplified on Fig. 7C, a non-monotonous relationship between local z coordinate and fluctuation amplitude was found. Two domains were apparent :
- for higher distances (z>70 nm), the fluctuation strongly increased as a function of distance, in accordance with predictions from the theory of statistical surfaces.
- when z became lower than about 50 nm, fluctuations did no longer decrease concomitantly. Interestingly, during the first tens of seconds following contact, low distance fluctuations increased, supporting the biological relevance of this finding.

iii) Examining deformation and fluctuation maps such as Fig.6B-C clearly revealed the heterogeneity of surface movement distribution, with a coexistence of more active and quieter membrane regions. The characteristic size of these areas was determined by studying spatial autocorrelation. A typical plot is displayed on Fig. 7D. The mean length for 50 % correlation decay determined on 16 separate cell images was 0.60 µm (0.45µm standard deviation). Accordingly, a comparable correlation length was found on the maps of individual pixel displacements during a fixed period of 2s, suggesting that cell membranes might deform as independent units of nearly micrometer size.

**Contact extension is correlated to important changes of cytoplasmic and surface molecules.**
In order to obtain more information on the structural basis of observed phenomena, fluorescence microscopy was associated with IRM to study the short term cell structural changes following contact formation

First, fluorescent antibodies were made to bind beta 1 integrins, i.e. the adhesion receptors mediating THP-1 cell attachment to fibronectin (24). High fluorescence levels were observed in cell-surface contact areas (not shown), but more extensive investigations are required to determine whether the surface density of integrins increased during contact maturation.

Second, we labelled CD43, a bulky membrane molecule that is highly abundant on leukocyte membranes and that is supposed to impair adhesive interactions (30, 31). Interestingly, as exemplified on Fig. 8, decrease of cell-to-substrate distance was associated with a concomitant egress of CD43 molecules.

Third, THP-1 cells were transfected with fluorescent actin and deposited on fibronectin-coated surfaces. Contact formation was very similar to processes described above, suggesting that transfection did not markedly alter THP-1 cell properties. The fluorescence distribution was then monitored together with IRM contacts. As exemplified on Fig. 9, in addition to bulk fluorescence, cell displayed localized fluorescence patches that seemed to appear a few tens of seconds before the formation or extension of of contact zones.

**Metabolic inhibitors influence contact extension.**
While it is well known that soft vesicles can display thermally-driven surface fluctuations and formation of extensive contacts with adhesive surfaces, it was important to determine to what extent cell behaviour was influenced by active processes dependent on biochemical events. This question was addressed as follows.
i) Stiffening cell surface with paraformaldehyde abolished contact formation and adhesion.
ii) Interestingly, treating cells with moderate concentrations of cytochalasin D (10 µg/ml) resulted in significant decrease of the rate of contact extension, smoothening of the two-dimensional contours of contact surfaces, which make them more alike contact zones formed with artificial vesicles (11,24), and increase of membrane undulations in membrane zones close to the adhesive surface (Fig. 10)
iii) Treating cells with a myosin inhibitor (butanedionemonoxime) to decrease cell motility abolished adhesion and contact formation.
iv) Since previous reports clearly demonstrated a role of intracellular calcium changes in cell spreading (32, 33), we investigated whether intracellular calcium level influenced THP-1cell fitting to adhesive surfaces. In preliminary experiments, cells were labelled with fluo-3, a fluorescent calcium probe, before deposition on fibronectin-coated surfaces and monitoring with IRM. While many cells displayed elevated calcium, no clear relationship was found between intracellular calcium and rate of contact growth (not shown).

**The requirement for fluid drainage from the cell-surface gap does not seem a limiting parameter for contact extension.**

In several situations of physical interest (e.g. aquaplaning) adhesion between surfaces may be hampered by intervening liquid film. The possibility that this might influence contact extension was investigated by monitoring contact extension in highly viscous ficoll solution. However, no significant decrease of contact extension was found when the medium viscosity was increased threefold (not shown).

## DISCUSSION

While cell attachment to adhesive surfaces and subsequent deformation and spreading were subjected to considerable scrutiny, little information is available on the mechanisms allowing transverse nanometer-scale adaptation of interacting surfaces. However, in view of the known roughness of cell membranes and length of adhesion molecules, this fitting process seems an obvious requirement to the formation of more than a few adhesive bonds. This point is important since older studies made on the strength of cell adhesion suggested that the minimal number of bonds required to account for experimental properties of cell adhesion was on the order of about at least one thousand (34, 35).

Interference reflection microscopy/reflection interference contrast microscopy was demonstrated to be a method of choice to image the fitting of cell membranes to planar surfaces with sufficient vertical resolution (18, 19, 22, 26). A major advantage of this technique is that it does not require any possibly artifactual treatment. It has been well recognized that the quantitative derivation of membrane-surface distance from intensity measurements relies on several assumptions that are difficult to prove (36) and the membrane-surface separation values we obtained may not be considered as absolute values of the distance between lipid bilayers and glass surfaces. However, it must be emphasized that the simple procedure we used to derive these distances is fairly robust and should provide reliable **relative values**, as acknowledged in previous papers (18, 20, 22). Indeed, cytoplasmic structures such as the actin cortex might influence image brightness. Although our observations of the kinetics of actin and distance map changes suggest that actin movements were much slower than the membrane fluctuations we described, a reliable proof of the validity of distance determination with IRM/RICM could only be obtained by comparing images obtained with different techniques. We are currently combining total internal reflection fluorescence microscopy and fluorescence labelling to achieve this goal.
Several important conclusions may be drawn from our experiments.

First, in addition to the mere confirmation of previous reports demonstrating the occurrence of transverse fluctuations of nucleated cell membranes (16-18), our results provide some new information on these fluctuations. The tentative view of these fluctuations as transverse motions of micrometer-size blocks with an effective diffusion coefficient of a few $nm^2/s$ may be useful to model these movements. Interestingly, this concept of independent motile units is consistent with older reports on the autonomous behaviour of small cytoplasmic fragments obtained from blood neutrophils (37). The patterns of cell undulations might also account for the recent findings that i) cell spreading on surfaces bearing adhesive structures requires a minimum surface density of anchoring points (38, 39), and ii) cell behaviour on adhesive surfaces is markedly influenced by topography (40, 41).

Second, the dependence of these fluctuations on local distance of the membrane to a surface (Fig. 7B) strongly suggests that cells perceive approaching structures at about 50 nm distance, resulting in increase of oscillations in contrast with an expected behaviour of passive objects (18, 29, 42). Thus, even the earliest period following cell encounter with a surface might involve active biological functions. Further work is needed to determine the surface molecules or the mechanisms mediating the initial interaction between cells and approaching surfaces.

Third, an important conclusion is that local membrane-to surface adhesion is not an instantaneous process, since the separation gap progressively decreases during the first minute following attachment (Fig. 5B). This phenomenon may be highly significant, since a recent study made on the initial adhesion between natural killer and target cells suggested that the width of the intercellular gap, as estimated with electron microscopy, might influence the outcome of adhesion (i.e. killer cell activation versus inactivation) (43). While little information is available on the dynamics of membrane tightening, it is reasonable to speculate that the egress of bulky membrane molecules such as CD43 might be a prerequisite for membrane approach. While the exit of CD43 (31, 44) and CD45 (45) from adhesion zones formed by immune cells is well demonstrated, the functional consequences remain incompletely understood. Indeed, these molecules may act either by influencing the presence of other signalling molecules, or they may by themselves trigger important biochemical reactions (such as phosphate removal by CD45).

Fourth, there remains to assess the mechanisms responsible for membrane fluctuations. Although the order of magnitude of these deformations is consistent with a passive thermal mechanism (18) as analyzed by Hellfrich (14), several possible mechanisms might allow active cell processes to influence these deformations, including i) removal of bulky membrane structures and reduction of repulsive forces, ii) ensuring low enough surface tension and bending rigidity, that are recognized as key parameters of passive deformations, and (iii) generation of active motion. Our findings that i) cytoskeletal impairment with cytochalasin D increased membrane fluctuations in regions closest to the adhesive surface, (ii) increased contact formation was often preceded by local actin concentration would be consistent with the hypothesis that microfilament might act by stabilizing the membrane and adhesive interactions. This interpretation would be in line with the old report (46) that detachment of surface adhesive cells might be triggered by a decrease of membrane associated microfilaments.

In conclusion, cell membrane fitting to adhesive surfaces is an early step of cell adhesion that occurs simultaneously with transverse membrane fluctuations. While strong evidence supports the hypothesis that these events are related by causal links, more work is required to unravel underlying molecular mechanisms.

**Acknowledgment**. This work was supported by a grant from the Association pour la Recherche sur le Cancer.

**FIGURE LEGENDS**

**FIGURE 1. Imaging contact formation between a cell and an adhesive surface.**
**A:** A spherical THP-1 cell deposited on a surface flattens within second as revealed with confocal microscopy in fluorescent medium. White bar length = 25 µm. **B:** the cell surface first appears fairly bright with interference reflection contrast microscopy. Bar length = 2 µm. **C:** Seventy-one seconds later an extensive contact zone was visible as a dark area with IRM/RICM. Bar length = 2 µm. **D:** Comparing the brightness histograms of images 1**B** (broken line) and 1**C** (full line) allows objective definition of contact zone with a cutoff intensity value (arrow). **E:** THP-1 cells were deposited on adhesive surfaces treated with different concentrations of fibronectin and the rate of contact growth was determined. Surfaces were derivatized with fibronectin solutions at 0.1 µg/ml (squares), 1 µg/ml (crosses), 10 µg/ml (triangles), 100 µg/ml (circles). Error bars represent standard errors of the mean of 15 to 25 separate cells.

**Figure 2. Typical pattern of growing contact area.**
**A-I:** Monocytic THP-1 cells were deposited on a surface treated with 10 µg/ml fibronectin. A microscope field was selected before the onset of the experiment and the cell-surface contact regions (corresponding to less than 34 nm apparent separation between surfaces) were determined at regular intervals of 9 seconds is shown as a black area. Bar length is 2 µm.

**Figure 3. Quantitative analysis of cell-surface separation.**
The distance between a cell and an adhesive surface was derived from IRM images and displayed with coded colours. **A**: cell membrane immediately after contact (time zero). **B** and **C**: images recorded at times 8s and 17s respectively. Bar = 2 µm.

**Figure 4. Three-dimensional image of the topography of the membrane zone facing an adhesive surface.**
The shape of the cell described on Fig. 3 is shown at times 0, 8, 17, 26, 35 and 64 s respectively on **A**, **B**, **C**, **D**, **E**, and **F**.

**Figure 5. Quantitative parameters of contact extension and tightening**.
**A:** The contact area determined on a single cell is shown, disclosing fluctuations that were not visible on mean curves displayed on Fig. 1**E**. **B:** The distance between a fixed point of the cell membrane (circle) and the minimum distance between the cell surface and the adhesive plane (crosses are shown), revealing progressive decrease of the distance between surfaces. This representative pattern was found with a similar time-scale on twelve different cells.

**Figure 6. Imaging cell membrane undulations.**
**A:** the contact evolution during a 5.6 second period of time is shown on a representative cell. While the region of stable contact is displayed in yellow, gain and loss of adhesive zone are displayed in red and green. **B:** The overall vertical membrane displacement is displayed with coded colours. **C** The amplitude of vertical membrane undulations is displayed as the standard deviation of eight sequential values of z coordinate, revealing the random part of the deformation.

**Figure 7. Quantitative description of cell membrane transverse fluctuations during contact extension.**
**A:** A representative cell imaged with about 1Hz frequency during contact formation. The squared displacement per pixel was plotted versus time after subtracting the overall displacement (corresponding to decrease of average cell-to-surface distance). The plot is fairly linear. Correlation coefficient is 0.991. **B:** Lifetime of membrane undulations: the time dependence of correlation between two sequential membrane maps (squares) on a typical cell reveals 50 % decrease of correlation during about 8 seconds. Crosses: displacements of a given membrane point reveal no detectable temporal correlation, suggesting that membrane deformations are driven by random movements with higher that 1 Hz frequency (in line with Fig. 3**B**). **C:** Dependence of the amplitude of vertical membrane displacements (as imaged in Fig. 6**C**) on distance to the adhesive surface. Squares: onset of contact. Crosses: 130 second after contact initiation. **D:** correlation length of cell contour. Both membrane images (squares) and local displacements during a 2.7 second interval (crosses) display marked spatial correlation with a 50 % decrease at 1 µm separation.

**Figure 8. Leukosialin redistribution during contact formation.**
Monocytic THP-1 cells were labelled with fluorescent anti-CD43/leukosialin before deposition on fibronectin-coated surfaces. Fluorescence and IRM images were then recorded at regular intervals. During the first minute, the separation between the cell membrane and the adhesive surface is nearly always higher than 35 nm (**A**), and CD43 is clearly visible on the whole surface, with local concentration (**B** - arrow). Sixty seconds later, cell-surface distance decreased by at least 20 nm (**C** - arrow) and CD43 depletion is clearly visible in the whole cell surface and particularly in the new contact zone (**D**).

**Figure 9. Actin redistribution during contact formation.**
Monocytic THP-1 cells were transfected with GFP before being deposited on fibronectin-coated surfaces and contact regions were analyzed with IRM and fluorescence. Two representative sets of images (**A-D** and **E-H**) are shown.
**A:** visible light image of a standard cell (with usual roundish appearance preceding spreading). **B** and **D** display contact growth during a 90 second period of time. On the fluorescence image (**C**), actin appears distributed throughout the whole cell, with local bright patches (arrow) that were associated with and seemed to precede contact.
**E-H**: a similar behaviour was observed several minutes after initial contact, on a markedly polarized cell.

**Figure 10. Effect of cytochalasin D on membrane fluctuations near a surface.**
Seven control cells (squares) and eight cytochalasin D-treated cells (crosses) were monitored for determination of membrane topographical and fluctuation maps. The standard deviation of membrane height, representing a fluctuation amplitude, was determined on separate groups of point with common distance to the surface. Cytochalasin treatment did not substantially influence fluctuations at 60-80 nm separation, but this seem to increase fluctuations for smaller distances (the difference was significant for 50 nm separation with student's t test).

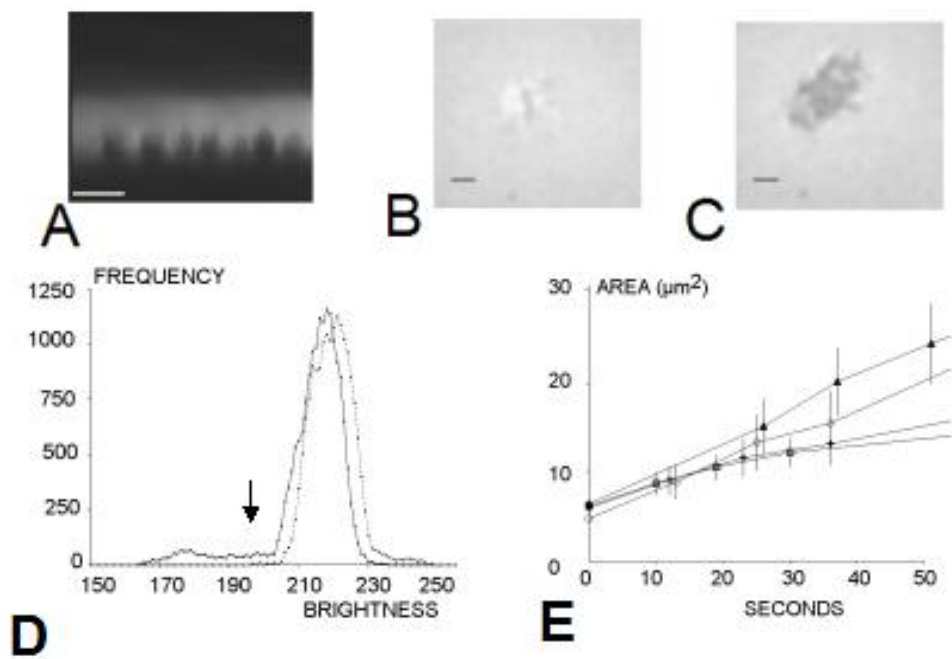

Fig. 1

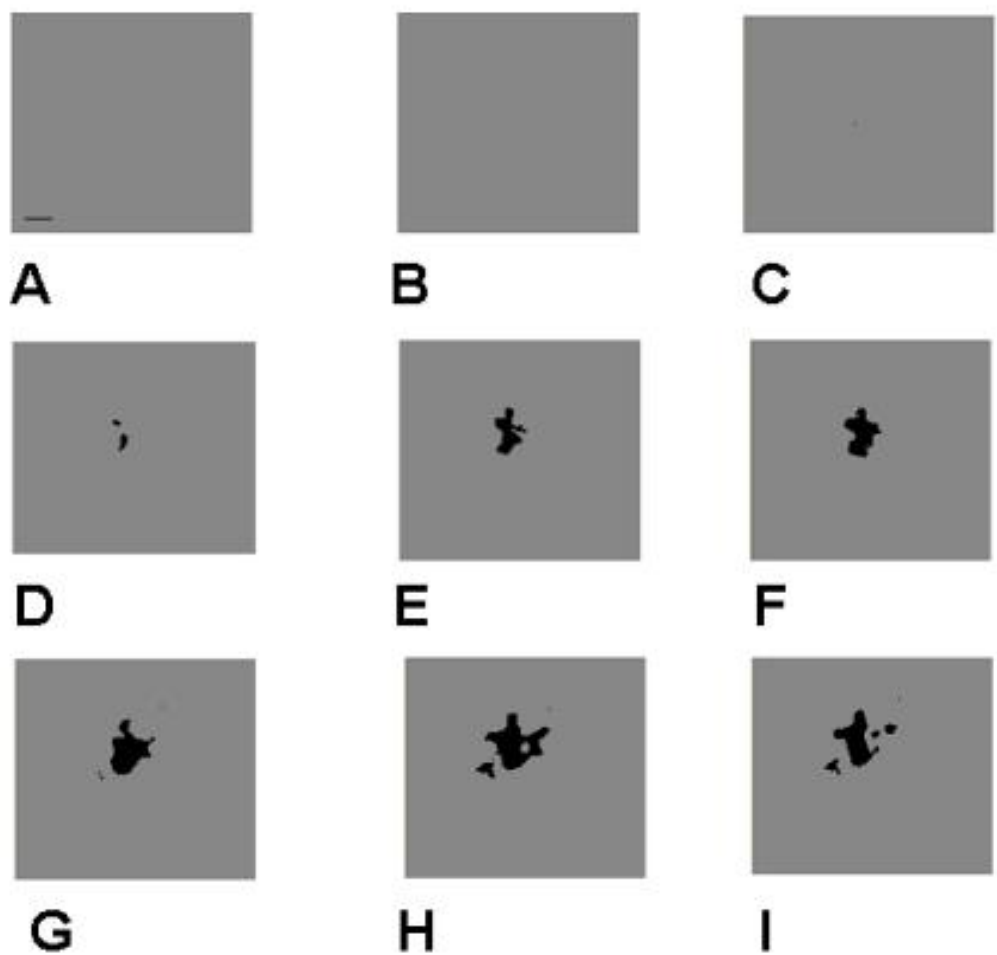

Fig. 2

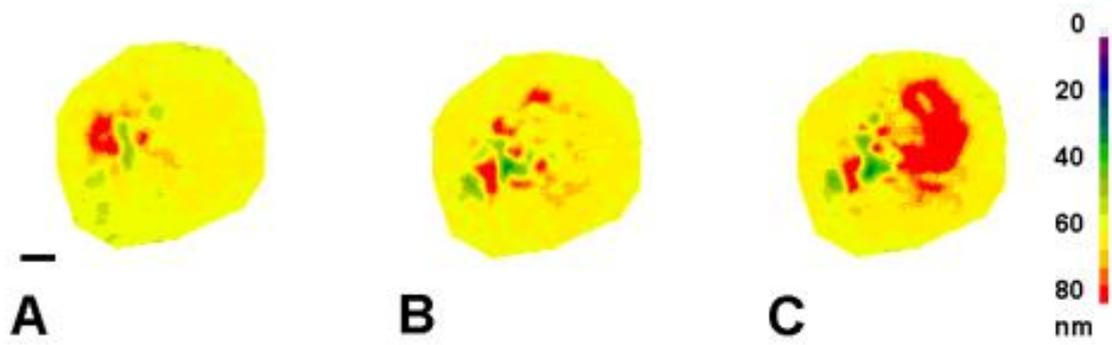

Figure 3

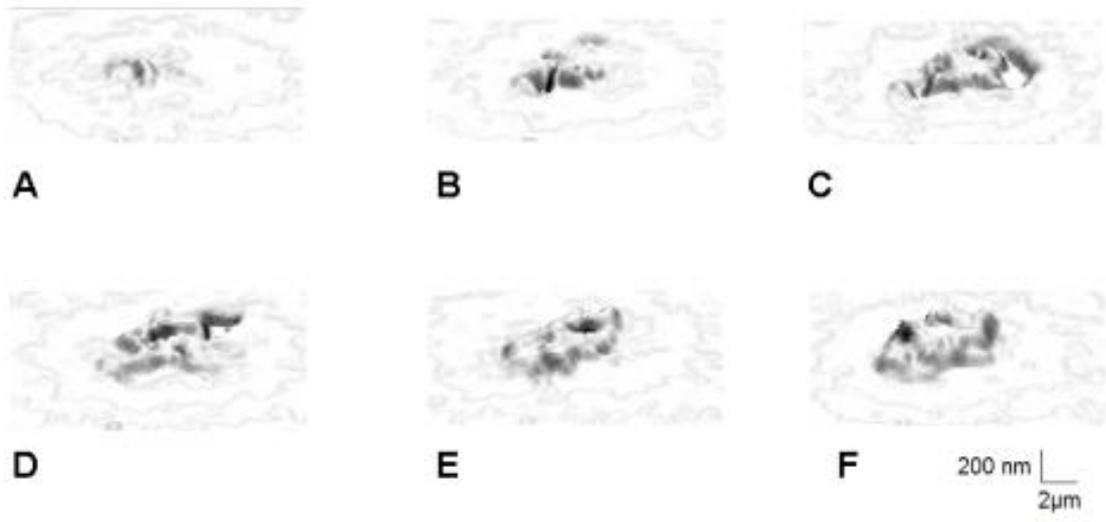

Fig. 4

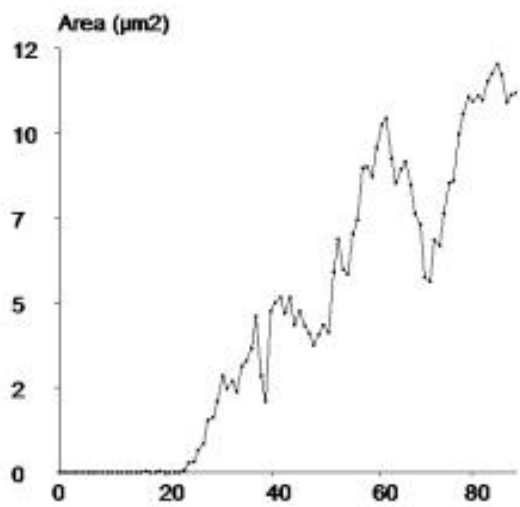 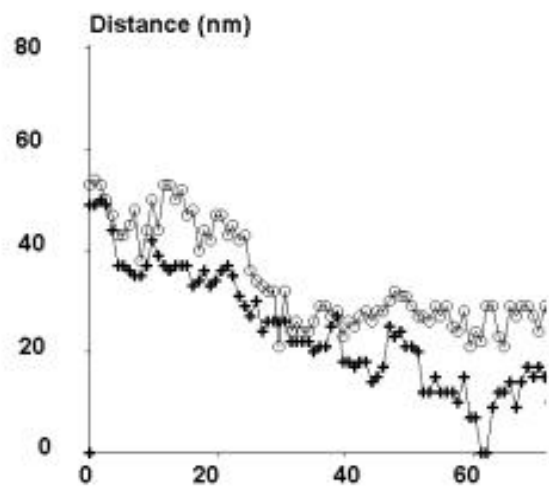

Fig. 5

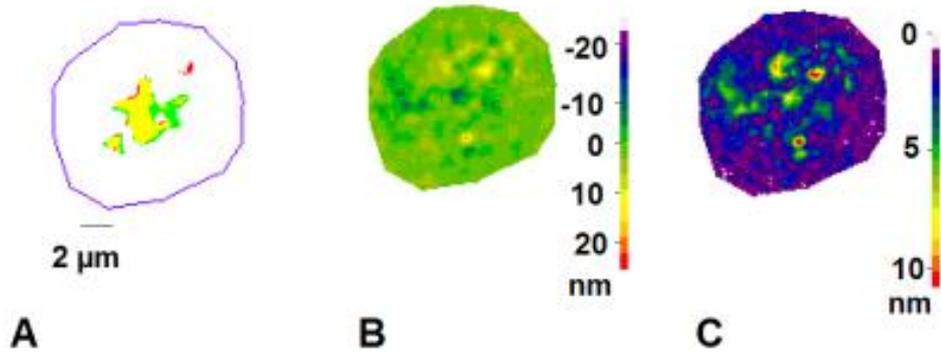

Fig. 6

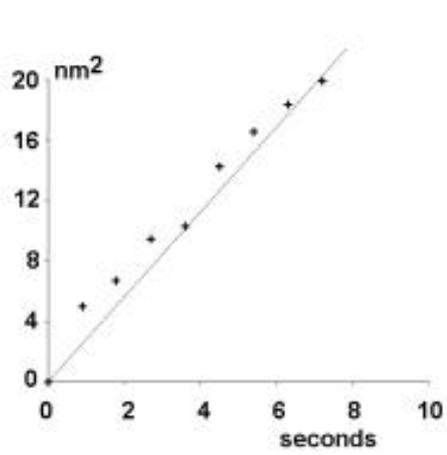 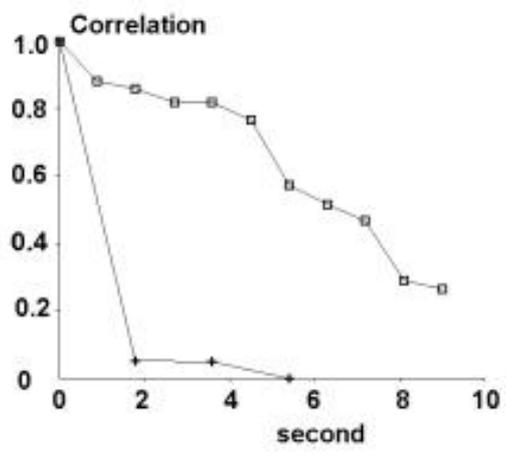
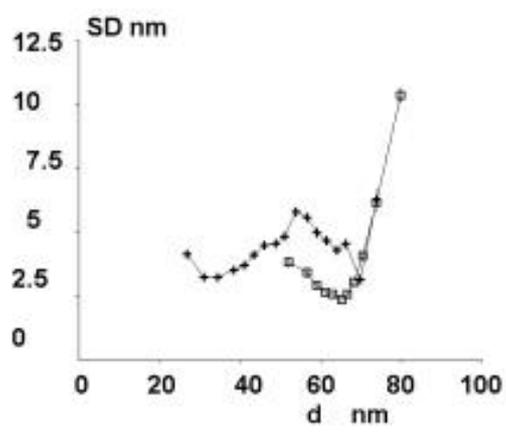 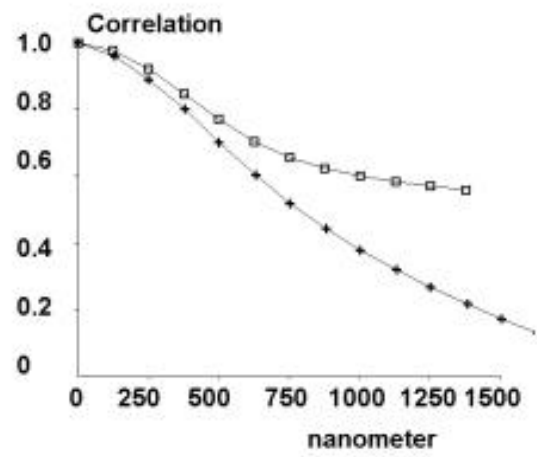

Fig. 7

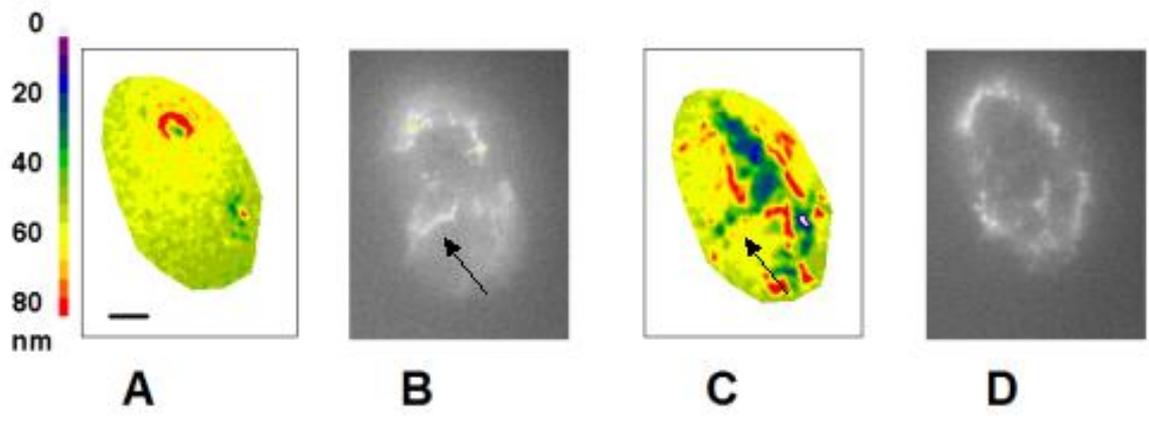

Fig. 8

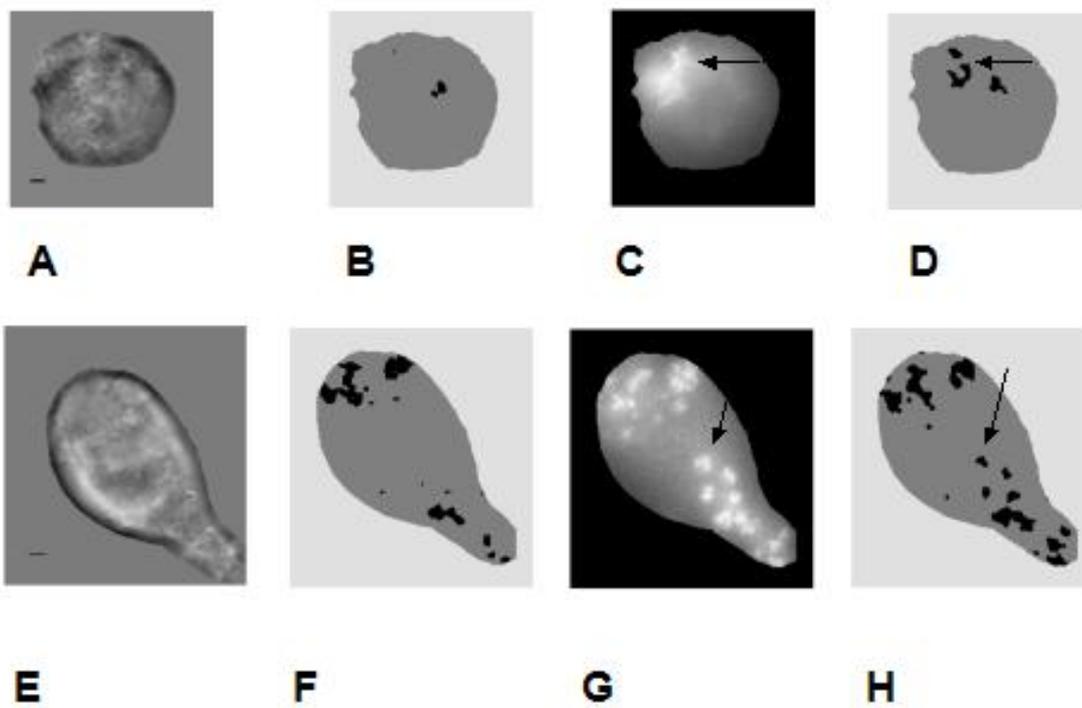

Fig. 9

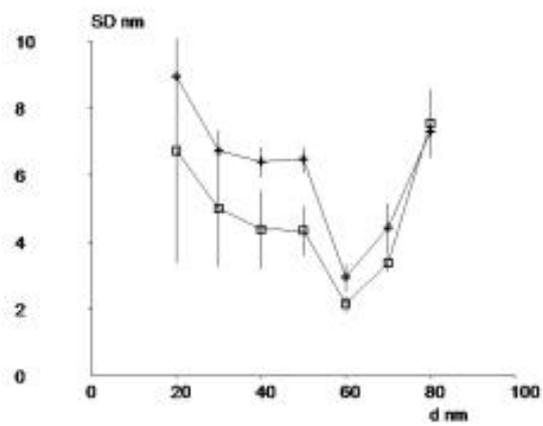

Fig. 10